\documentclass[fleqn,usenatbib,useAMS]{mnras}

\usepackage[T1]{fontenc}
\usepackage{xspace}
\usepackage{xcolor}
\usepackage[detect-all]{siunitx}

\DeclareRobustCommand{\VAN}[3]{#2}
\let\VANthebibliography\thebibliography
\def\thebibliography{\DeclareRobustCommand{\VAN}[3]{##3}\VANthebibliography}

\usepackage{graphicx}	% Including figure files
\usepackage{amsmath}	% Advanced maths commands
\usepackage{amssymb}	% Extra maths symbols
\usepackage{bookmark}
\usepackage{nicefrac}

\newcommand{\pd}{\textit{P(D)}\xspace}
\newcommand{\pofdaffine}{\texttt{pofd\_affine}\xspace}
\newcommand{\hspire}{\textit{Herschel}-SPIRE\xspace}
\newcommand{\herschel}{\textit{Herschel}\xspace}
\newcommand{\planck}{\textit{Planck}\xspace}

\usepackage{newtxtext,newtxmath}

\title[\textit{Herschel} Dark Field P(D) Fluctuation Analysis]{The \textit{Herschel}-SPIRE Dark Field II: A \textit{P(D)} Fluctuation Analysis of the Deepest \textit{Herschel} Image of the Submillimetre Universe}

\author[T. W. O. Varnish et al.]{
Thomas W. O. Varnish,$^{1,2,3}$\thanks{E-mail: \href{mailto:tvarnish@mit.edu}{tvarnish@mit.edu}}
Xinni Wu,$^{1}$
Chris Pearson,$^{2, 4, 5}$
David L. Clements$^{1}$ 
and Ayushi Parmar$^{1}$
\\
$^{1}$Imperial College London, Blackett Laboratory, Prince Consort Road, London SW7 2AZ, UK\\
$^{2}$RAL Space, UKRI STFC Rutherford Appleton Laboratory, Chilton, Didcot, Oxfordshire OX11 0QX, UK\\
$^{3}$Plasma Science and Fusion Center, Massachusetts Institute of Technology, Cambridge, MA 02139, USA\\
$^{4}$Department of Physical Sciences, The Open University, Milton Keynes, MK7 6AA, UK\\
$^{5}$Oxford Astrophysics, Denys Wilkinson Building, University of Oxford, Keble Road, Oxford, OX1 3RH, UK\\
}

\date{Accepted February 10, 2025. Received January 10, 2025; in original form April 9, 2024.}

\pubyear{2025}

\begin{document}
\label{firstpage}
\pagerange{\pageref{firstpage}--\pageref{lastpage}}
\maketitle

\begin{abstract}
The \hspire Dark Field is the deepest field produced by the SPIRE instrument pushing down below the galaxy confusion limit in each of the 250, 350, 500 $\umu$m bands. Standard source extraction techniques inevitably fail because of this, and we must turn to statistical methods. Here, we present a P(D)---probability of deflection---analysis of a 12$'$ diameter region of uniform coverage at the centre of the \hspire Dark Field. Comparing the distribution of pixel fluxes from our observations to the distributions predicted by current literature models, we find that none of the most recent models can accurately recreate our observations. Using a P(D) analysis, we produce a fitted differential source count spline with a bump in the source counts at faint flux densities, followed by a turnover at fainter fluxes, required to fit the observations. This indicates a possible missing component from the current literature models, that could be interpreted perhaps as a new population of galaxies, or a missing aspect of galaxy evolution. Taking our best-fit results, we also calculate the contribution to the CIB in each of the bands, which all agree with the \planck CIB measurements in this field.
\end{abstract}


%%%%%%%%%%%%%%%%%%%%%%%%%%%%%%%%%%%%%%%%%%%%%%%%%%

%%%%%%%%%%%%%%%%% BODY OF PAPER %%%%%%%%%%%%%%%%%%

\section{Introduction} \label{sec:intro}
% \textcolor{red}{(1) Introduce the CIB, and confusion. (2) Brief recap of Herschel. (3) Previous attempts to find dN/dS: direct counts, statistical methods, stacking, etc. (4) What have we done?}
% INTRODUCTION. Let's add some text to the thingy.

% \begin{itemize}
	% \item Why should we care about all this shizzle. Half of the emission from galaxies is in the far-infrared: therefore, studying these wavelengths is an important step.
% \end{itemize}

% \begin{itemize}
% 	\item Herschel
% 	\item SPIRE
% 	\item Prior campaigns
% 	\item DSFGs + CIB
% 	\item attempts
% 	\item New dark field
% 	\item overview --> statistical approach
% 	\item Summarize what you're about to read
% \end{itemize}

% \textcolor{red}{Introduce dust, CIB, etc. more generally. Throw in a few citations?}
In 1967, a background of light emitted from young galaxies was predicted by \citet{Partridge1967a, Partridge1967b}. Subsequent surveys in the optical to identify the galaxies responsible for this background were unsuccessful \citep{Davis1974}. Ultimately, it was discovered that interstellar dust played a key role in obscuring the background's origin, absorbing and reprocessing the ultraviolet (UV) light emitted from young stellar objects to far-infrared (FIR) wavelengths \citep{Kaufman1976}, meaning this background could lie between 30-500 $\umu$m \citep{Stecker1977}. Detection of this predicted cosmic infrared background (CIB) was ultimately made with the FIRAS instrument \citep{Mather1993, Puget1996, Fixsen1998} aboard the COBE satellite \citep{Boggess1992}, and later confirmed by DIRBE \citep{Schlegel1998}. Infrared observations are important for understanding galaxy evolution; the energy contained within the CIB is comparable with the energy contained within the cosmic optical background (COB) emission, i.e. approximately half of the emission from galaxies is at far-infrared wavelengths. As such, it follows that to fully understand the evolution of galaxies we should also understand their emission at FIR wavelengths, not just their emission in optical bands \citep{Ryter1977, Soifer1987}. 

Launched in 2009, the \textit{Herschel} Space Observatory provided a revolutionary step forward in exploring the cosmic history of dust-obscured galaxies and large-scale structure formation \citep{Pilbratt2010}. The SPIRE instrument on-board \textit{Herschel}, in particular, was able to make simultaneous observations at wavelengths of 250, 350, and 500 $\umu$m \citep{Griffin2010}. The \textit{Herschel} mission carried out many legacy programmes over its lifespan, including the HerMES \citep{Oliver2012}, H-ATLAS \citep{Eales2010, Clements2010}, and HLS \citep{Egami2010} surveys. Of particular importance to this study is the \textit{Herschel} Lensing Survey (HLS). However, all deep \textit{Herschel} surveys (except the HLS survey that used gravitational lensing of clusters to reach the deepest fields) suffered from confusion noise due to the relatively low resolution of the \hspire instrument at far-infrared (FIR) wavelengths, and the high number density of FIR sources on the sky. During \textit{Herschel}'s four-year mission, the \hspire instrument routinely observed a patch of dark sky for calibration of the instrument \citep{Pearson2014}. By combining 141 of these observations, we have assembled the deepest \hspire image ever produced \citep{Pearson2025}: roughly 10 times the integration time of the HLS survey.

\cite{Pearson2025} presented source count results from conventional source extraction (SUSSEXtractor) and list-driven cross-identification (XID) methods \citep{Hurley2017} on the \hspire Dark Field, revealing a good agreement with literature counts at fluxes $\gtrsim$20 mJy. With conventional source extraction methods (SUSSEXtractor), our results at fainter fluxes ($\lesssim$20 mJy) were significantly affected by confusion noise. We, therefore, applied a list-driven XID source extraction technique---using 24 $\umu$m catalogues from the Spitzer-MIPS instrument as a prior---to begin to probe around and below the confusion-limited flux level.

However, to probe to even fainter fluxes ($\lesssim$1 mJy), we must turn to statistical methods, such as P(D), to exploit the unique depth of the \hspire Dark Field \citep{Scheuer1957}. Previous work on \textit{Herschel} data has already shown this technique to be a valuable method of overcoming the significant confusion noise seen in \hspire observations \citep{Glenn2010}.

In Section \ref{sec:obs}, we describe the data and observations of the \hspire Dark Field used in this study. In Section \ref{sec:pd}, we present the P(D) method and results from this analysis of the \hspire Dark Field. In Section \ref{sec:cib}, we use the results from our P(D) analysis to calculate the contribution of these sources to the CIB and compare the results to \planck HFI observations. In Section \ref{sec:discussion} we discuss the potential interpretations of these new P(D) differential source count results. We finally conclude by proposing potential extensions and implications of these results in Section \ref{sec:conclusion}.

\section{Observations} \label{sec:obs}
In this study, we make use of observations of the \hspire Dark Field \citep{Pearson2025}. The \hspire Dark Field is composed of 141 stacked, mean-subtracted, calibration observations in scan-map mode of a roughly circular patch of sky, 30$'$ in diameter, near the North Ecliptic Pole at R.A. = 17h40m12s, Dec. = +69d00m00s. It is the deepest field observed by the SPIRE instrument. These observations reach below the source confusion limits in every SPIRE band: PSW (250 $\umu$m), PMW (350 $\umu$m), and PLW (500 $\umu$m). Because of this, the \hspire Dark Field is highly confused, which makes it ideally suited for a statistical P(D) approach once standard source extraction techniques are exhausted.

Within the \hspire Dark Field, we define a smaller region at the centre of the maps which we use for our analysis. We refer to this 12$'$ diameter, circular region as the Deep Region. This smaller field is selected as it has an almost uniform coverage and is void of any contamination from significant diffuse emission or bright sources. When using a statistical technique like P(D) analysis, it is important to select a uniform field so as not to affect the statistical interpretation of the pixel counts. Non-uniformities in the coverage will affect the interpretation of the pixel brightnesses.

The instrumental noise for the (250, 350, 500) $\umu$m SPIRE band observations has been measured to scale as (9.0, 7.5, 10.8) mJy per $\sqrt{N_{reps}}$, where $N_{reps}$ is the total number of map repeats in the observation \citep{Griffin2010}. The SPIRE Dark Field is assembled from 141 maps, each with 4 repeats (for a total of 564 repeated observations). Table \ref{tab:noise} presents the calculated instrument noise levels for the observations used in our study. In our maps, we expect scaled instrumental noise levels of (0.379, 0.316, 0.455) mJy in the (250, 350, 500) $\umu$m bands respectively.

Generally speaking, for an image containing a total of $N_{pix}$ pixels, we might assume the fluxes at each pixel is an independent measurement. However, as \cite{Scheuer1957} notes, this assumption is invalid for confused fields with low angular resolutions. Instead, since the SPIRE beam is larger than a single pixel, the emission from a point source will be spread over multiple pixels. As a result, there is a non-zero correlation between pixel intensities. So, our standard error (or limiting sensitivity of the P(D) technique) becomes $\sigma_{inst}/\sqrt{N_{beams}}$. Since a P(D) analysis considers the statistical distribution of pixel intensities (see Section \ref{sec:pd}) rather than attempting to identify and extract sources, we are able to reach below the confusion limit, and the instrument noise, all the way down to the limiting sensitivity. Table \ref{tab:noise} presents the calculated limiting sensitivity for each SPIRE band.

%%%%%%%%%%%%%%%%%%%%%
% TABLE: Instrument noise, limiting sensitivity, and confusion noise.
\begin{table}
	\centering
	\caption{Instrument noise ($\sigma_{\text{inst}}$), limiting sensitivity ($\sigma_{\text{limit}}$), and confusion noise ($\sigma_{\text{conf}}$), for each SPIRE band. The confusion noise here is reproduced from \protect\cite{Nguyen2010}. The instrument noise for each band is from \protect\cite{Griffin2010}, scaled by the square root of the number of repetitions in each band in the Dark Field. All noises and limits are given in mJy.}
	\label{tab:noise}
	\begin{tabular}{lccc}
		\hline
		SPIRE Band & $\sigma_{\text{inst}}$ [mJy] & $\sigma_{\text{limit}}$ [mJy] & $\sigma_{\text{conf}}$ [mJy] \\
		\hline
		250 $\umu$m (PSW) & 0.379 & 0.00643 & 5.8$\pm$0.3 \\
		350 $\umu$m (PMW) & 0.316 & 0.00714 & 6.3$\pm$0.4 \\
		500 $\umu$m (PLW) & 0.455 & 0.0151 & 6.8$\pm$0.4 \\
		\hline
	\end{tabular}
\end{table}
%%%%%%%%%%%%%%%%%%%%%

\section{P(D) Analysis} \label{sec:pd}
To extend previous galaxy survey work done on the Deep Field \citep{Pearson2025}, we have conducted a P(D)---probability of deflection---analysis on the region. A P(D) analysis considers the probability distribution of flux deflections above/below the mean flux level in an observation, due to an underlying population of faint-flux sources \citep{Scheuer1957}. This distribution is built from a given source count model, and the point spread function of the band/instrument in question \citep{Barcons1990}. We can therefore vary this input source count model to match the observed distribution of flux deflections as seen in the Deep Field observations. This technique has already been used for \hspire observations by \cite{Glenn2010} to great effect, but on a significantly shallower field.

\subsection{P(D) Method} \label{sec:pd:method}
Following the method of \citet{Glenn2010}, we make use of the \pofdaffine package\footnote{The \pofdaffine package was developed by A. Conley and is available at \url{https://github.com/aconley/pofd_affine}} for our P(D) analyses. This software represents the source count models as splines. These splines are made up of knots, where each knot defines a Euclidean normalised differential source count value [Jy$^{1.5}$ deg$^{-2}$] (the knot value) at a specified flux [mJy] (the knot flux). The spline interpolation smoothly connects these knots.

The \pofdaffine package uses a Markov Chain Monte Carlo (MCMC) method to optimise the source count model to match a given observed flux deflection distribution. This flux deflection distribution is calculated by taking a histogram of the pixel intensity values from an observation image.

We initially constrain the values of any knots---where the knot fluxes $\gtrsim2$ mJy---using the \cite{Glenn2010} results. 
For knots with fainter fluxes $\lesssim2$ mJy, where there are no longer any observed number count results, the initial source count values of the knots were chosen manually to follow the trends predicted by the current literature models \citep{Negrello2017, Pearson2025}. 
These knot values were then allowed to evolve from these initial conditions, according to the MCMC optimisation routine used by the \pofdaffine package.
On the order of ${\sim}100$ optimisations were run with these different input conditions to ensure that our optimisation had found the global optimal solution, and not just local minima.

To assess the goodness-of-fit of our P(D) results, we calculate the reduced chi-squared statistic for the predicted pixel flux distribution and the observed pixel flux distribution. Additionally, we compared our best-fit results with the results predicted by current literature models \cite{Bethermin2017, Negrello2017, Pearson2025}. These comparisons are presented in Figure \ref{fig:pd_fit_allBands_negrello} and Table \ref{tab:chisq_results}.

We estimate the uncertainty in our fitted source count spline result using the following procedure. Taking a copy of the best-fit spline, randomised displacements in the knot value are applied to each knot. The reduced chi-squared (RCS) statistic is recalculated for this perturbed spline. If the new spline's RCS value lies within $1\sigma$ of the best-fit spline's RCS statistic, we record the knot values for each knot flux. Starting again from a new copy of the best-fit spline, this process is repeated until we have accumulated enough accepted splines that we can take statistics of the distribution of the possible values for each knot. Our estimated $1\sigma$ uncertainty region is then given by the mean and standard deviation of each distribution of knot fluxes for each knot and interpolated linearly between fluxes in log space since we treat each knot independently in this procedure.

% To find the uncertainties in our fitted source count models, we take the best-fit \textcolor{red}{spline} determined by \pofdaffine, and within a specified region surrounding this \textcolor{red}{spline}, randomly vary \textcolor{red}{the source count value of each knot}. For each new model with a set of randomly varied knots, if the reduced chi-squared value indicates a good fit (i.e. the reduced chi-squared value is less than the best-fit reduced chi-squared plus 1$\sigma_{\text{RCS}}$), we record the knot values. Ultimately, this produces a distribution of possible knot values for each knot. Using these distributions, we define 1$\sigma$ and 2$\sigma$ regions around each knot. However, since the knots are often strongly correlated, these uncertainty regions don't always overlap with every single knot on the best-fit curve. This is most noticeable in our PSW results.

\subsection{P(D) Results} \label{sec:pd:results}
\begin{figure*}
\begin{minipage}{\textwidth}
	\centering
	\includegraphics[width=\textwidth]{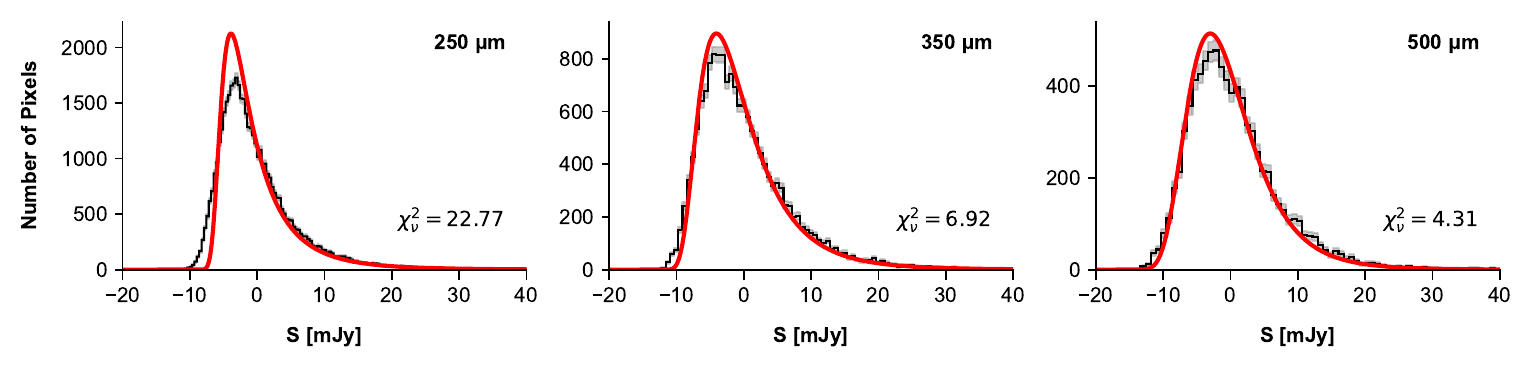}
	\includegraphics[width=\textwidth]{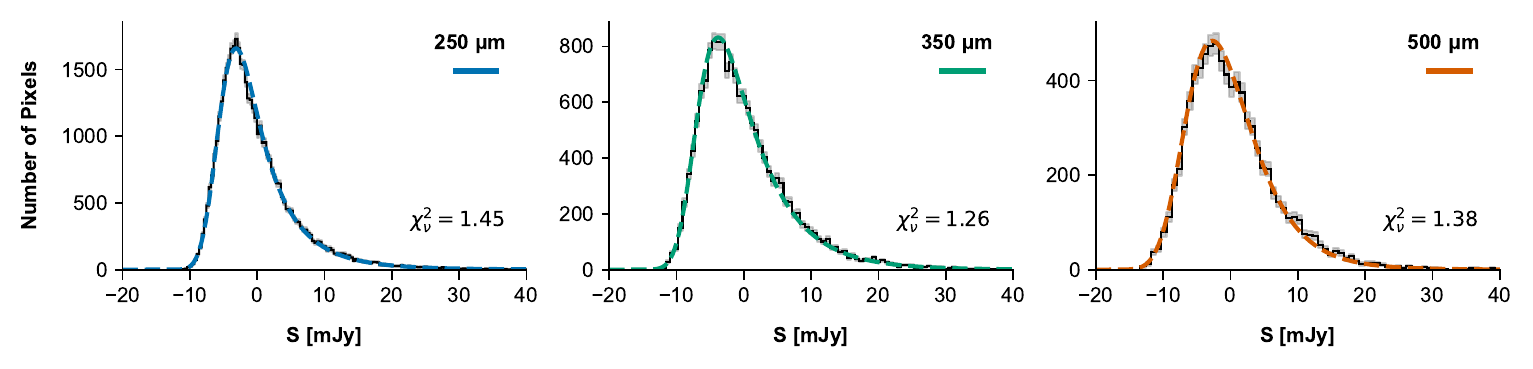}
	\vspace{-5mm}
	\caption{P(D) results. Top row: comparison between observed pixel flux distribution (black histogram) with flux distribution predicted by the \citep{Negrello2017} model (red curve). Bottom row: comparison with the best-fit splines presented in this paper.} \label{fig:pd_fit_allBands_negrello}
\end{minipage}
\end{figure*}

Figure \ref{fig:pd_fit_allBands_negrello} presents comparisons between the observed flux distributions from the Deep Region of the \hspire Dark Field with the predicted flux deflection distributions from our P(D) analysis, and the predictions of the \citet{Negrello2017} galaxy source count model. In all three bands, the predicted distribution from the \citet{Negrello2017} source count model struggles to fit the observed fluctuations at faint fluxes. Therefore, to fit our observed pixel distributions these models will need modifying. Following the method outlined in Section \ref{sec:pd:method}, we produced best-fit P(D) splines for the Dark Field. The predicted distributions from our best-fit P(D) splines, as seen in Figure \ref{fig:pd_fit_allBands_negrello}, match the observed pixel distributions (unlike the literature models). These qualitative conclusions are supported by the relevant reduced chi-squared values for these fits (seen alongside the plots in Figure \ref{fig:pd_fit_allBands_negrello}, and in Table \ref{tab:chisq_results}).

%%%% FIGURE: P(D) results. %%%%%
%%%%%%%%%%%%%%%%%%%%%%%%%%%%%%%%

Table \ref{tab:chisq_results} presents the reduced chi-squared statistics (goodness-of-fit) for our best-fit source count splines from our P(D) analysis, along with comparisons to three current source count models from the literature. A well-fitted model will have a reduced chi-squared statistic, $\chi_{\nu}^2$, close to 1. To represent an overall goodness-of-fit considering all three SPIRE bands, we compare the average $\chi_{\nu}^2$ for each model. As is clear from the reduced chi-square statistics and the visual comparison with the observed pixel flux distribution, none of the current literature source count models can accurately describe our observations of the \hspire Dark Field. We compare: 
\begin{itemize}
	\item \citet{Bethermin2017}: a phenomenological model, based on earlier work, with dark-matter halos included to better match observed clustering
	\item \citet{Negrello2017}: extends an earlier model by \citet{Cai2013} (which includes only late (spirals) and early (spheroids) populations), by including a population of lensed spheroids.
	\item \citet{Pearson2025}: a backwards evolution infrared-based model which includes quiescent, starburst, luminous, ultraluminous, and AGN populations.
\end{itemize}

%%%% TABLE: P(D) fits RCS %%%%%%
\begin{table}
	\centering
	\caption{Reduced chi-squared statistics are presented for three literature models for each of the three SPIRE bands. None of the current literature models are able to accurately describe the \textit{Herschel}-SPIRE Dark Field \pd results. Sorted in order of descending Avg. $\chi_\nu^2$}
	\label{tab:chisq_results}
	\begin{tabular}{lcccc}
		\hline
		& \multicolumn{4}{c}{$\chi_{\nu}^{2}$} \\
		\multicolumn{1}{l}{Model} & 250 $\umu$m & 350 $\umu$m & 500 $\umu$m & Avg. \\
		\hline
		% \citet{Pearson2019} & & 206.82 & 147.75 & 83.26 & 145.94 \\
		\citet{Bethermin2017} & 18.44 & 15.99 & 16.00 & 16.81 \\
		\citet{Negrello2017} & 22.77 & 6.92 & 4.31 & 11.33 \\
		\citet{Pearson2025} & 16.17 & 7.11 & 4.49 & 9.26 \\
		\hline
		Our P(D) Analysis & 1.45 & 1.26 & 1.38 & 1.36 \\
		\hline
	\end{tabular}
\end{table}
%%%%%%%%%%%%%%%%%%%%%%%%%%%%%%%%

We generated expected flux deflection histograms for each of the three literature models above. Comparing these histograms with the observed flux deflection histograms (see Figure \ref{fig:pd_fit_allBands_negrello} for an example), we see that none of the models can properly describe our Deep Field observations. Quantitatively, we can see that the reduced chi-squared statistics for these fits agree that the models cannot describe the data. Table \ref{tab:chisq_results} presents these goodness-of-fit results.

\begin{figure}
	\includegraphics[width=\columnwidth]{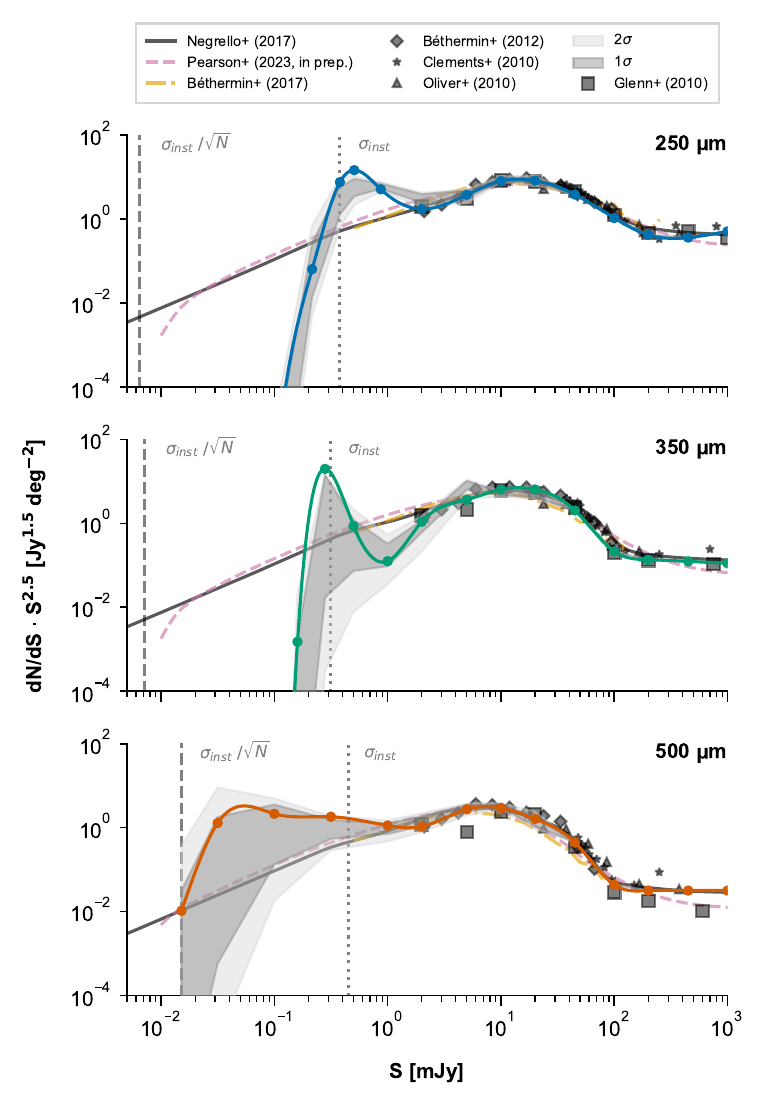}
	\vspace{-5mm}
    \caption{Differential source counts for all three SPIRE bands (250, 350, 500) $\umu$m. Observed differential source count results (black points) show very good agreement at bright fluxes. \citep{Clements2010, Oliver2010, Bethermin2012}. Prior SPIRE \pd results are also presented \citep[black squares]{Glenn2010}, and demonstrate our field is capable of reaching much fainter fluxes. Our \pd results (grey regions) show a bump-turnover feature going towards fainter fluxes. Our \pd results deviate from current literature galaxy evolution models (solid, dashed, and dotted lines) at these fainter fluxes \citep{Bethermin2017, Negrello2017, Pearson2019}. We indicate the instrument noise ($\sigma_{\text{inst}}$) and limiting sensitivity ($\sigma_{\text{inst}} / \sqrt{N}$) fluxes for reference (grey dotted and dashed lines, respectively). }
    \label{fig:bestfit_dNdS}
\end{figure}

Figure \ref{fig:bestfit_dNdS} presents our best-fit source count splines and regions of 1$\sigma$ and 2$\sigma$ uncertainty (grey shaded regions) resulting from our P(D) analysis. Table \ref{tab:bestfit_dNdS} lists the values of the plotted spline knots. In all three SPIRE bands, we see a good agreement with the previously observed source counts down to ${\sim}$20 mJy \citep{Clements2010, Oliver2010}. We also continue to see a good agreement at fainter fluxes to the stacked counts from \citet{Bethermin2012}, and also the P(D) counts from \citet{Glenn2010}, down to ${\sim}$2 mJy. At fluxes $\gtrsim$2 mJy, we also see good agreement with the source count models from the literature \citep{Bethermin2017, Negrello2017, Pearson2025}. 

However, at fluxes $\lesssim$2 mJy---below the limit of previously observed counts---our P(D) results begin to deviate from the trend predicted by the literature models. Going to fainter fluxes, as we approach the instrument noise limit in all three bands we see a significant increase in our P(D) source count results compared with the literature models. In the 350 $\umu$m band, we also see a significant dip in the source counts (at ${\sim}$1 mJy) preceding this steep rise. In all three bands, the literature models do not predict this trend.

% Below the instrument noise level---since we should be able to draw some statistical conclusions about our data down to the limiting sensitivity defined in Section \ref{sec:obs}
Below the instrument noise level (Section \ref{sec:obs}), we see a steep drop off in the Euclidean normalised source counts as we go to fainter fluxes in the 250- and 350-$\umu$m bands. The resulting ``bump-turnover'' shape in the source counts peaks at ${\sim}0.5$ mJy in the 250 $\umu$m band and at ${\sim}0.3$ mJy in the 350 $\umu$m band. The turnover seen is also much more sudden than the gentle roll-off predicted by the literature models. In previous, complementary work on the CIB, \cite{Duivenvoorden2020} were similarly able to push below the instrument noise level using a stacking technique.

In the PLW band, we see a similar albeit slightly different trend below the instrument noise level. Instead of a sharp peak near the instrumental noise limit, we see a gentle rise above the predictions given by the literature models, leading into a roughly steady Euclidean normalised source count that doesn't change with decreasing flux. This trend peaks slightly at ${\sim}0.05$ mJy, before turning over and beginning a rapid decrease in source counts with decreasing flux. We are not able to measure the full turnover due to the cut-off imposed by the limiting sensitivity. This turnover in the counts in the PLW band happens at much lower (by around an order of magnitude) fluxes than in the PSW and PMW bands. The peaks in the source count models observed in each band occur at lower fluxes with increasing wavelength. We will discuss the potential interpretations of these results in Section \ref{sec:discussion}.

In each band, we see our uncertainty regions follow the general trend of the best-fit spline. The region also widens as we go to fainter fluxes. Strangely, however, our best-fit spline deviates from the uncertainty region, most notably in the PSW band. We hypothesise this is due to how we have treated the spline knots when estimating the uncertainty region. The knots in the best-fit spline are highly correlated and must represent a coherent spline whereas the points defining the uncertainty region are treated independently. These uncertainty regions are ranges in number counts where a spline could lie which does not contradict with the observations at that sigma level. For our best-fit spline, this coincides with the point not quite matching in the histogram and suggests that there may be more structure in the number counts than we can reproduce. It is important to emphasise that our best-fit spline is not a rigorous physical model, but a fit to the data.

Another possible source of uncertainty could be in the mean-subtraction performed on the Dark Field maps. If the mean of the map is incorrect, this will add an offset to all pixel flux values in the map, and could affect the differential counts. An uncertainty of 0.07 MJy/sr at 250 microns corresponds to an offset of approximately 0.5 mJy/beam. Adjusting the mean by this value could potentially move the counts from the peak in the best-fit spline at 0.4 mJy to the lowest flux knot. However, considering the pixel histograms in Figure \ref{fig:pd_fit_allBands_negrello}, a mean offset would equate to a lateral shift of the histogram in flux, $S$. As we can see, with the current literature models we still cannot explain our observations by simply shifting the distributions in $S$. We still require new models to produce a different histogram shape overall. This can be quantified by comparing the reduced chi-squared statistics for maps with/without an offset applied to the mean. For our best-fit PSW spline, applying a shift of $\pm 0.5$mJy/beam increases the reduced chi-squared statistic for our fit from ${\sim}1.5$ (with no mean shift) to ${\sim}6.5$ (with the shift applied). Table \ref{tab:shifts} summarizes the results for the \cite{Negrello2017} model in all three bands. In all cases, adding a shift to the mean produces a higher reduced chi-squared statistic, and as such a less good fit to the observations.

\begin{table}
	\centering
	\caption{Quantitive goodness-of-fit tests on whether a shift applied to the mean could explain the poorly-fitting literature models. For each row, a constant offset---$\pm 0.5$ mJy/beam---is applied to the maps before the histogram is calculated and compared to the literature model presented in Fig. \ref{fig:pd_fit_allBands_negrello} \citep{Negrello2017}.}
	\label{tab:shifts}
	\begin{tabular}{lccc}
	    \hline
     & \multicolumn{3}{c}{Reduced Chi-Squared, $\chi_\nu^2$} \\
      Mean Shift [mJy/beam] & 250 $\umu$m & 350 $\umu$m & 500 $\umu$m \\
      \hline
      -0.5 & 30.30 & 10.97 & 5.66 \\
      0.0 & 22.77 & 6.92 & 4.31 \\
      +0.5 & 23.78 & 7.77 & 6.70 \\
     \hline
	\end{tabular}
\end{table}

%%%%%%%% FIGURE: BEST-FIT source count MODELS %%%%%%%%%%%%%
%%%%%%%%%%%%%%%%%%%%%%%%%%%%%%%%%

%%%%%%%%% TABLE %%%%%%%%%%%%%%%%%
\begin{table*}
    \begin{minipage}{\textwidth}
	\centering
	\caption{Euclidean normalised best-fit and 1$\sigma$ differential source count results from our P(D) analyses in all three \hspire bands. The best-fit results (along with the 1$\sigma$ uncertainty region bounded by the 1$\sigma$ lower and upper values) are plotted in Figure \ref{fig:bestfit_dNdS}.}
	\label{tab:bestfit_dNdS}
	\begin{tabular}{cccc|cccc|cccc}
		\hline
		\multicolumn{4}{c|}{250 $\umu$m} & \multicolumn{4}{c|}{350 $\umu$m} & \multicolumn{4}{c}{500 $\umu$m} \\
        \hline
        & Best-Fit & \multicolumn{2}{c|}{$1\sigma$} & & Best-Fit & \multicolumn{2}{c|}{$1\sigma$} & & Best-Fit & \multicolumn{2}{c}{$1\sigma$} \\
		Flux & $\nicefrac{\text{dN}}{\text{dS}} \cdot S^{2.5}$ & Lower & Upper & Flux & $\nicefrac{\text{dN}}{\text{dS}} \cdot S^{2.5}$ & Lower & Upper & Flux & $\nicefrac{\text{dN}}{\text{dS}} \cdot S^{2.5}$ & Lower & Upper \\
        $[\log_{10} (\text{Jy})]$ & \multicolumn{3}{c|}{$[\log_{10} (\text{Jy}^{1.5} \text{deg}^{-2})]$} & $[\log_{10} (\text{Jy})]$ & \multicolumn{3}{c|}{$[\log_{10} (\text{Jy}^{1.5} \text{deg}^{-2})]$} & $[\log_{10} (\text{Jy})]$ & \multicolumn{3}{c}{$[\log_{10} (\text{Jy}^{1.5} \text{deg}^{-2})]$} \\ 
		\hline
		-4.30 &   -17.0 &  -20.7 &  -16.5 &       -4.00 &   -15.0 &  -26.8 &  -16.8 &       -4.82 &  -1.98 &  -6.90 &  -1.50 \\
        -4.00 &   -6.09 &  -8.22 &  -5.65 &       -3.80 &   -2.82 &  -8.13 &  -2.74 &       -4.50 &  0.110 &  -3.24 &  0.256 \\
        -3.67 &   -1.19 &  -1.87 & -0.575 &       -3.55 &    1.30 &  -1.77 &   1.16 &       -4.00 &  0.330 & -0.882 &  0.552 \\
        -3.42 &   0.884 &  0.115 &  0.772 &       -3.30 & -0.0623 &  -1.13 & 0.0868 &       -3.50 &  0.260 & -0.266 &  0.185 \\
        -3.29 &    1.17 &  0.547 &  0.962 &       -3.00 &  -0.900 &  -1.02 & -0.483 &       -3.00 & 0.0500 & -0.152 & 0.0976 \\
        -3.06 &   0.712 &  0.723 &  0.848 &       -2.70 &  0.0400 & -0.216 & 0.0659 &       -2.70 & 0.0250 & 0.0408 &  0.268 \\
        -2.70 &   0.241 &  0.396 &  0.596 &       -2.30 &   0.567 &  0.805 &   1.01 &       -2.30 &  0.440 &  0.384 &  0.543 \\
        -2.30 &   0.584 &  0.483 &  0.684 &       -2.00 &   0.810 &  0.679 &  0.747 &       -2.00 &  0.470 &  0.480 &  0.564 \\
        -2.00 &   0.911 &  0.907 &  0.996 &       -1.70 &   0.809 &  0.616 &  0.807 &       -1.70 &  0.209 & 0.0396 &  0.254 \\
        -1.70 &   0.918 &  0.868 &  0.996 &       -1.35 &   0.311 &  0.311 &  0.311 &       -1.35 & -0.353 & -0.353 & -0.353 \\
        -1.35 &   0.588 &  0.588 &  0.588 &       -1.00 &  -0.668 & -0.668 & -0.668 &       -1.00 & - 1.36 &  -1.36 &  -1.36 \\
        -1.00 &  0.0268 & 0.0268 & 0.0268 &      -0.699 &  -0.864 & -0.864 & -0.864 &      -0.699 &  -1.50 &  -1.50 &  -1.50 \\
        -0.699 & -0.375 & -0.375 & -0.375 &      -0.347 &  -0.906 & -0.906 & -0.906 &      -0.347 &  -1.50 &  -1.50 &  -1.50 \\
        -0.347 & -0.439 & -0.439 & -0.439 &       0.000 &  -0.950 & -0.950 & -0.950 &       0.000 &  -1.50 &  -1.50 &  -1.50 \\
         0.000 & -0.290 & -0.290 & -0.290 &           - &       - &      - &      - &           - &      - &      - &      - \\
		\hline
	\end{tabular}
    \end{minipage}
\end{table*}
%%%%%%%%%%%%%%%%%%%%%%%%%%%%%%%%%

% \begin{itemize}
%         \item Our new model (from fitting) has a bump-turnover shape. Bump: more emission than expected by models. Turnover: all sources detected? (Do we discuss this here, or leave it until after the CIB section?)
% \end{itemize}

\section{Contribution to the CIB} \label{sec:cib}
The emission from a large population of galaxies contributes to a cosmic (infrared) background. Using our best-fit $dN/dS$ results from our P(D) analysis, we can integrate the $dN/dS$ curve to obtain a contribution to the extra-galactic background intensity from our population,
\begin{equation}
    \int S \cdot \frac{dN}{dS} \cdot dS,
\end{equation}
where $dN/dS$ are the differential (not Euclidean normalised) source counts. Integrating over the entire flux range finds the total emission from the entire population of faint galaxies. Comparing this with independent measurements of the CIB then enables us to verify the results of our P(D) analysis. With \pofdaffine, it is possible to constrain the fit using a known integrated CIB value. However, we chose not to do this so that we could independently verify the result using observations of the same region from the \planck satellite.

\subsection{Comparison with Planck Observations}
\begin{figure}
	\includegraphics[width=\columnwidth]{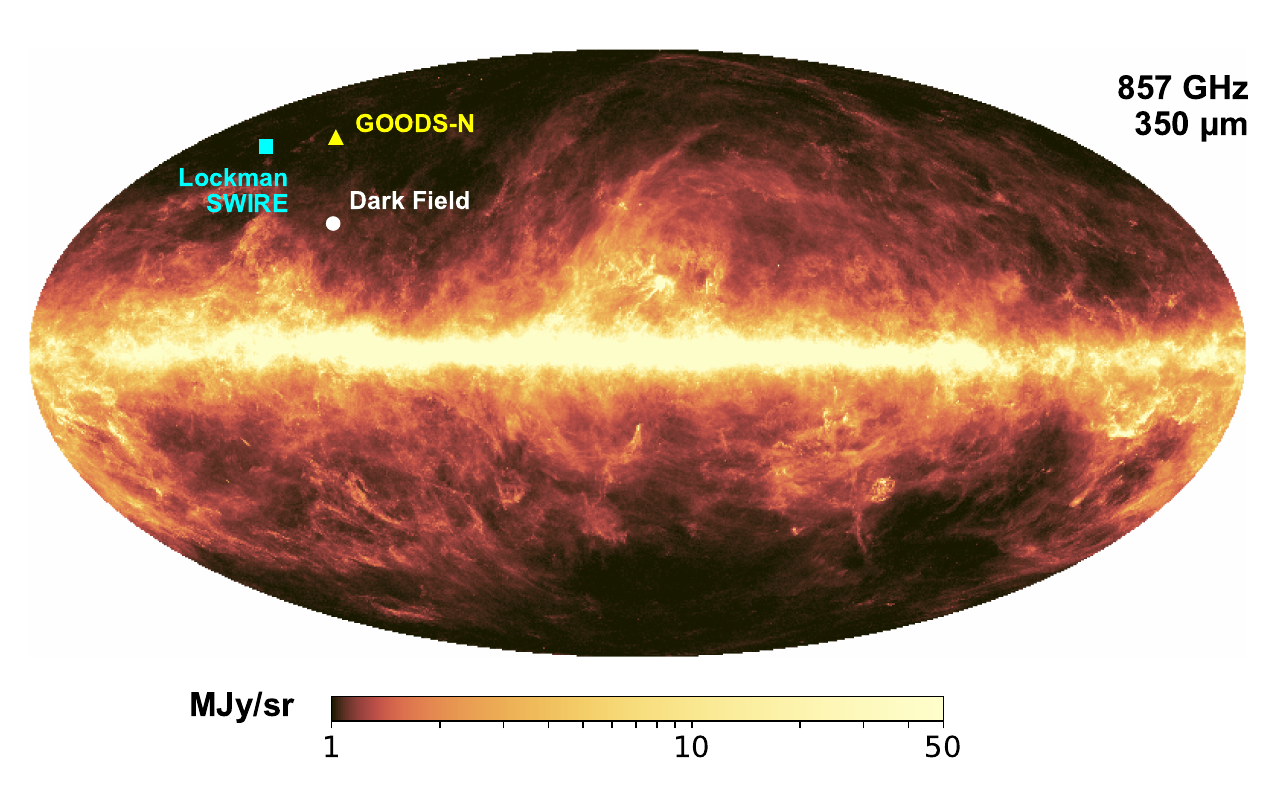}
	\vspace{-5mm}
    \caption{An all-sky map, in galactic coordinates, at 857 GHz (or 350 $\umu$ m) from the \planck satellite. Locations of various survey fields (including the Dark Field) are indicated. Lockman-SWIRE (cyan square), GOODS-N (yellow triangle), and the \hspire Dark field (white circle).}
    \label{fig:planck}
\end{figure}

The High Frequency Instrument (HFI) aboard \planck captured all-sky maps\footnote{The \planck all-sky maps can be found at \url{https://irsa.ipac.caltech.edu/data/Planck/release_3/all-sky-maps/}} at 545 GHz (550 $\umu$m) and 857 GHz (350 $\umu$m) \citep{Ade2014}.
Figure \ref{fig:planck} shows the 857 GHz (350 $\umu$m) map with the locations of the Dark Field region and other SPIRE fields indicated.
Using these maps, we cut out the region corresponding to the Deep Region within the \hspire Dark Field, and integrated the \planck measurements taken within this region. This gave us a total intensity in MJy/sr for the region of sky in which our P(D) $dN/dS$ models were fitted.

In order to compare these observations with our \hspire data, we converted the 545 GHz (550 $\umu$m) maps to 500 $\umu$m by assuming the intensity ratio of the CIB as measured by \planck at these two wavelengths was the same as the ratio of the CIB as measured by the FIRAS instrument aboard COBE at these two wavelengths \citep{Fixsen1998}. This is in essence a colour correction. At a wavelength of $\lambda$ $\umu$m, we measure an intensity of $I_{F,\lambda}$ and $I_{P,\lambda}$ with COBE-FIRAS and \planck, respectively. So, we have:
\begin{equation}
    \frac{I_{F,550}}{I_{F,500}} = \frac{I_{P,550}}{I_{P,500}} \longrightarrow I_{P, 500} = \frac{I_{P, 550} \cdot I_{F, 500}}{I_{F, 550}}
\end{equation}
and using this we can calculate the intensity at 500 $\umu$m, as would be measured by the \planck and \herschel satellites. Similarly, we also colour-corrected the \planck results from $350$ $\umu$m to $250$ $\umu$m, to enable a direct comparison with the \hspire PSW results.

% \begin{itemize}
%     \item High Frequency Instrument (HFI) all-sky maps at 545 GHz (550 $\umu$m) and 857 GHz (350 $\umu$m).
%     \item Converted 545 GHz maps to 500 $\umu$m using: (double-check technique).
% \end{itemize}

\subsection{CIB Results}
Table \ref{tab:CIB_results} presents the results from our integration of both the SPIRE P(D) fit and the \planck data. This includes the $250{-}$ and $500{-}\umu$m colour-corrected \planck data, as described in Section \ref{sec:cib}.

All three of the \planck measurements lie within the $1\sigma$ region bounded by our SPIRE results. However, we get a closer match between the SPIRE best-fit result ($2.20\pm0.07$) and the \planck result ($2.011\pm0.069$) in the 350 $\umu$m band. According to our SPIRE best-fit results, the deep region is redder (more emission in the longer-wavelength bands) than what was measured by \planck.

\begin{table}
	\centering
	\caption{Integrated CIB results for each SPIRE band: PSW (250 $\umu$m), PMW (350 $\umu$m), and PLW (500 $\umu$m). We have the integrated CIB from our best-fit source count model for each band and for the upper and lower 1$\sigma$ uncertainty regions. We take the bounding $dN/dS$ curve for this uncertainty region and integrate it to find the lower and upper bounds. We also present an integrated CIB result from the \planck satellite of the Deep Region. These independent measurements lie within our 1$\sigma$ bounds for the SPIRE results, and we see the best agreement in the 350 $\umu$m band.}
	\label{tab:CIB_results}
	\begin{tabular}{lccc}
	    \hline
     & \multicolumn{3}{c}{Intensity [MJy/sr] in SPIRE Band} \\
      & 250 $\umu$m & 350 $\umu$m & 500 $\umu$m \\
      \hline
    1$\sigma$ Lower-Bound & 1.56$\pm$0.05 & 0.53$\pm$0.02 & 0.54$\pm$0.02 \\
    Best-Fit & 2.27$\pm$0.07 & 2.20$\pm$0.07 & 2.59$\pm$0.08 \\
    1$\sigma$ Upper-Bound & 2.73$\pm$0.08 & 2.63$\pm$0.08 & 4.0$\pm$0.1 \\
     \hline
     \planck (Deep Region) & 2.6$\pm$0.6
 & 2.01$\pm$0.07 & 0.98 $\pm$ 0.11 \\
     % PLANCK (Lockman-SWIRE) & - & 0.875\$\
    \hline
	\end{tabular}
\end{table}

\section{Discussion} \label{sec:discussion}

Without pushing to fainter fluxes, our results agree well with the current literature models for fluxes $\gtrsim 20$ mJy. However, it is the unique depth of the \hspire Deep Region---that we can exploit with statistical techniques like P(D)---that has enabled us to explore fainter fluxes than ever before with the \hspire instrument. At these fainter fluxes $\lesssim 20$ mJy, the bump-then-turnover in the source counts models found via the P(D) analysis indicates our observations account for 100\% of the FIR galaxies at 250 and 350 $\umu$m. \citet{Duivenvoorden2020} applied a novel source extraction technique to the COSMOS field \citep{Scoville2007} and found results consistent with recent literature galaxy count models and the \citet{Fixsen1998} CIB measurements. The CIB convergence and agreement with the \citet{Fixsen1998} result suggests that \cite{Duivenvoorden2020} have observed all the galaxies within the COSMOS field between 250 $\umu$m and 500 $\umu$m.

Our results are also---at faint ($\lesssim 20$ mJy) fluxes---distinct from the current literature models. The bump-turnover shape, which cannot be explained by any of the current models, indicates that something is missing from these models. This could be an additional contribution from a new population of galaxies, or a population that is already known but simply absent in the models. Perhaps it could be an evolution mechanism that is not included or poorly understood. With the data we have, we are unable to conclude exactly what this missing component is. Further observations or modelling is necessary to discover this, but this is beyond the scope of this paper.

We also notice that the peak of the bump in the source counts shifts to fainter fluxes as the wavelength increases. If this bump is due to an unaccounted-for population in the literature models, this shifting towards fainter fluxes with increasing wavelength suggests the population might be bluer in colour (brighter at shorter wavelengths).

As with the \citet{Duivenvoorden2020} results, the close agreement of our 350 $\umu$m CIB result with that from the \planck satellite is yet another indication (in addition to the sharp turnover in the source counts) that we have observed all the FIR galaxies at this wavelength within the \hspire Deep Region.

Although the \planck measurement of the CIB at 500 $\umu$m lies comfortably within the region bounded by the lower and upper $1\sigma$ bounds, our SPIRE measurement from integrating the best-fit source count spline predicts a value almost two and a half times larger. It is important to note that the uncertainty region for our 500 $\umu$m spline result (Fig. \ref{fig:bestfit_dNdS}) is the largest of all three bands.

% \begin{itemize}
%     \item Instead of using a catalogue of sources to find contribution to CIB, we can take distribution of sources (dN/dS) and integrate:
%     $$ a $$
%     to find total contribution to CIB from model. With the turnover, we've detected all the sources we can, so this integrated CIB should converge!
%     \item Duivenvoorden did similar but with catalogue from stacking analysis. Our approach is statistical, but still ascribes all the emission to galaxies!
% \end{itemize}

% \begin{itemize}
%     \item Turnover in P(D) source counts model indicates our map accounts for 100\% of the FIR galaxies at 250/350 microns. \textcolor{red}{Never before seen directly in the source count data? But what does \citep{Duivenvoorden2020} say?} Duivenvoorden: new fitting technique to get catalogue. Use catalogue to find CIB from these sources. They agree with models! Find more precise than FIRAS. Suggest CIB contribution converged.
%     \item Distinct from current models; includes bump at fainter fluxes which isn't seen. Suggests additional contribution unaccounted for by models. Is this a new population? Or is it a new evolution mechanism? We can't conclude.
%     \item Peak in bump shifts to lower fluxes at longer wavelengths --> bluer? Foreground galaxies?
%     \item Turnover indicates we've accounted for 100\% of sources. Both 250 and 350 $\umu$m clearly show turnover. 500 $\umu$m begins to turnover but not as dramatically as in PSW/PMW. Even uncertainty regions indicate need for this bump/turnover shape.
% \end{itemize}

\section{Conclusions} \label{sec:conclusion}
% SUMMARY OF CIB

% DEEP REGION IN DARK FIELD

Using data from the \hspire Dark Field (constructed from stacked calibration maps), we identified a region of uniform coverage at the centre we call the Deep Region \citep{Pearson2025}. This region is highly confused due to the large beam size of the \hspire instrument, and as such, standard source extraction techniques are limited. Instead, we performed a statistical P(D) analysis of the region \citep{Scheuer1957, Glenn2010}. A P(D) analysis attempts to fit a distribution of flux deflections due to an underlying population of faint-flux sources (galaxies).

Comparing current literature models with our best-fit $dN/dS$ source counts from the P(D) analysis, we found that none of the current models were able to describe our data, particularly at faint ($\lesssim 20$ mJy) fluxes. In all three bands, our best-fit spline had a bump-turnover shape going to lower fluxes. A bump---above the trend in the source counts predicted by current literature models---was observed at ${\sim}0.5$ mJy, ${\sim}0.3$ mJy, and ${\sim}0.05$ mJy in the 250 $\umu$m, 350 $\umu$m, and 500 $\umu$m bands respectively. Both the 250 $\umu$m and 350 $\umu$m bands exhibited a rapid turnover in the source counts, indicating that we had observed the entire population of galaxies in those two bands with our observations. In the 500 $\umu$m band, we still see a turnover, albeit at much fainter fluxes, and less steep than in the other two shorter-wavelength bands.

Integrating our best-fit source counts splines, we also saw the CIB contribution agrees with independent \planck observations at 350 $\umu$m. At 500 $\umu$m, the \planck emission lies within the $1\sigma$ bounds of our \hspire measurement, although it agrees less closely. Our 500 $\umu$m CIB measurements are less certain, likely due to most of the emission being at such faint fluxes ($\lesssim 0.1$ mJy) and therefore our uncertainty on these values is greater. However, the turnover and agreement with the CIB in the 350 $\umu$m band does suggest that we've detected the entire 350 $\umu$m population of galaxies in this region with our observations. 

The Deep Region overlaps deep MIPS and IRAC observations in the Spitzer IRAC validation field \citep{Krick2008}, and while this already provides a wealth of ancillary data, extending the observations of the Deep Region to submillimetre wavelengths via ground-based facilities may help to further constrain the fainter population through higher resolutions. The \hspire Dark Field observations presented here are the deepest far-IR observations currently available, and they will remain so until the next generation of space-based far-IR telescopes start to operate. Such telescopes will be necessary for us to better understand the nature and origin of the sub-mJy bump in counts found by our P(D) analysis. The ideal instrument for this task should be both more sensitive than \hspire and also capable of operating at a higher angular resolution so that it is not as confusion-limited in the far-IR. Several far-IR mission concepts are currently being studied by NASA to fulfil the far-IR Probe recommendations of the US Decadal Review \citep{DecadalSurvey}, most of which will be able to achieve some of these capabilities either directly, or through the use of ancillary data at other wavelengths to break confusion. Until such a future mission flies we are limited to using data from \hspire, the deepest of which, as discussed here, is found in the \hspire Dark Field.

% \begin{itemize}
%     \item Using stacked calibration maps (Herschel-SPIRE Dark Field), we're looking at the deepest-ever Herschel image with a statistical technique. Previous work \cite{Pearson2025} has looked at this field with standard source extraction techniques and agrees with bright-flux models. Our new work pushes these results to fainter fluxes.
%     \item None of the current literature models are able to explain our data. We fit a new source counts model that agrees better (compare reduced chi-squared values, perhaps) than any of the current models.
%     \item Our fitted model requires a bump in the source counts just below the previous SPIRE fields, followed by a turnover in the counts towards fainter fluxes.
%     \item Turnover in the counts suggests our model accounts for the entire population of galaxies in each band.
%     \item CIB contribution shown to agree in PMW. Planck emission lies within bounds for PLW, but CIB from PLW is very uncertain (likely due to such faint fluxes, and therefore uncertainty).
%     \item PMW turnover and agreement with CIB does suggest that we've detected and ascribed sources to the complete CIB emission in this band.
%     \item \textcolor{red}{TODO: future study ideas.}
% \end{itemize}

\section*{Acknowledgements}

% SPIRE OBSERVATIONS
SPIRE has been developed by a consortium of institutes
led by Cardiff University (UK) and including: University of Lethbridge
(Canada); NAOC (China); CEA, LAM (France); IFSI, University of 
Padua (Italy); IAC (Spain); Stockholm Observatory (Sweden); 
Imperial College London, RAL, UCL-MSSL, UKATC,
University of Sussex (UK); and Caltech, JPL, NHSC, University of Colorado (USA). 
This development has been supported by national funding agencies: 
CSA (Canada); NAOC (China); CEA, CNES, CNRS (France); ASI 
(Italy); MCINN (Spain); SNSB (Sweden); STFC, UKSA (UK); 
and NASA (USA). HIPE is a joint development by the \textit{Herschel} 
Science Ground Segment Consortium, consisting of ESA, the NASA \textit{Herschel}
Science Center, and the HIFI, PACS and SPIRE consortia.

% SPITZER OBSERVATIONS
Our work also made use of observations from the Spitzer Space
Telescope, which is operated by the Jet Propulsion Laboratory, 
California Institute of Technology, under a contract with the 
National Aeronautics and Space Administration (NASA). 

Our research made use of the NASA/IPAC Infrared Science Archive, which is 
funded by the National Aeronautics and Space Administration and 
operated by the California Institute of Technology
Funding for this work was provided, in part, by STFC.

% Imperial HPC
We acknowledge computational resources and support provided by the Imperial College Research Computing Service (\url{http://doi.org/10.14469/hpc/2232}), and support via the RAL Space In House Research programme funded by the Science and Technology Facilities Council of the UK Research and Innovation (award ST/M001083/1).

% Colourmaps
Our paper makes use of perceptually uniform, colour-vision-deficiency-friendly colour maps \citep{Crameri2023} to prevent the visual distortion of the data and promote accessibility.

The authors would also like to thank the anonymous referee for their comments and suggested improvements to the paper.

%%%%%%%%%%%%%%%%%%%%%%%%%%%%%%%%%%%%%%%%%%%%%%%%%%
\section*{Data Availability}
The processed \hspire Dark Field observations (including the Deep Region) are available on Zenodo at \url{https://dx.doi.org/10.5281/zenodo.14846645}, and are discussed in more detail in \cite{Pearson2025}. P(D) and $dN/dS$ model files, along with plotting scripts (python jupyter notebooks) for the data analysed in this study are also available on Zenodo under the same directory. The best-fitting source count splines for all three SPIRE bands (250, 350, 500 $\umu$m) are presented in Tables \ref{tab:bestfit_dNdS}, and are available in full online. Datasets for the CIB analysis were derived from sources in the public domain. The \planck all-sky maps are accessible at \url{https://irsa.ipac.caltech.edu/data/Planck/release_3/all-sky-maps/}. The IRAS ISSA all-sky maps can be found at \url{https://lambda.gsfc.nasa.gov/product/iras/iras_thirdp_prod_table.cfm}



\bibliographystyle{mnras}
\bibliography{references} % if your bibtex file is called example.bib

\bsp	% typesetting comment
\label{lastpage}
\end{document}